\documentclass[prb,twocolumn,showpacs,preprintnumbers,amsmath,amssymb]{revtex4}

\usepackage{graphicx}
\usepackage{dcolumn}
\usepackage{bm}

\begin{document}

\preprint{APS/123-QED}

\title{High-energy spectroscopic study of the III-V nitride-based\\
diluted magnetic semiconductor Ga$_{1-x}$Mn$_{x}$N}

\author{J. I. Hwang$^{1}$, Y. Ishida$^{1}$, M. Kobayashi$^{1}$, H. Hirata$^{1}$, K. Takubo$^{1}$, T. Mizokawa$^{1}$ and A. Fujimori$^{1,2}$,\\
J. Okamoto$^{2}$, K. Mamiya$^{2}$, Y. Saito$^{2}$, Y. Muramatsu$^{2}$, H. Ott$^{3}$, A. Tanaka$^{4}$, T. Kondo$^{5}$ and H. Munekata$^{5}$}
\affiliation{
$^{1}$Department of Complexity Science and Engineering and Department
of Physics,\ University of Tokyo,\ Kashiwanoha,\
Kashiwa-shi,\ Chiba,\ 277-8561,\ Japan\\
$^{2}$Synchrotron Radiation Research Center,\ Japan Atomic Energy Research
Institute,\ SPring-8,\ Kouto,\ Mikazuki,\ Sayo-gun,\ Hyogo 679-5148,\ Japan\\
$^{3}$Institut f\"ur Experimentalphysik Freie Universit\"at Berlin,\
Arnimallee 14 D-14195 Berlin,\ Germany\\
$^{4}$Department of Quantum Matter,\ ADSM,\ Hiroshima University,\ Kagamiyama,\
Higashi-Hiroshima-shi,\ Hiroshima 739-8530,\ Japan\\
$^{5}$Imaging Science and Engineering Laboratory,\ Tokyo Institute of Technology,\ Nagatsuda,\ Midori-ku,\ Yokohama-shi,\ 226-8502,\ Japan}

\date{\today}

\begin{abstract}
We have studied the electronic structure of the diluted magnetic
semiconductor Ga$_{1-x}$Mn$_{x}$N ($x$ = 0.0, 0.02 and 0.042) 
grown on Sn-doped $n$-type GaN using 
photoemission and soft x-ray
absorption spectroscopy. Mn $L$-edge x-ray absorption have indicated 
that the Mn ions are in the tetrahedral crystal field and that their valence 
is divalent. 
Upon Mn doping into GaN, new state were found to form
within the band gap of GaN, and the Fermi level was
shifted downward. Satellite structures in the Mn 2$p$ core level
and the Mn 3$d$ partial density of states were analyzed using
configuration-interaction calculation on a MnN$_{4}$ cluster model.
The deduced electronic structure parameters reveal that the $p$-$d$ exchange coupling
in Ga$_{1-x}$Mn$_{x}$N
is stronger than that in Ga$_{1-x}$Mn$_{x}$As. 
\end{abstract}

\pacs{75.50.Pp,\ 75.30.Hx,\ 79.60.-i,\ 78.70.Dm}
\maketitle
\section{Introduction}
Diluted magnetic semiconductors (DMS) have recently attracted much
interest because of novel functions and potential applications
of the combination of magnetism caused by the local spins of
transition-metal ions and semiconducting properties due to the
itinerant carriers of the host materials. In the II-VI and III-V
DMS such as Cd$_{1-x}$Mn$_{x}$Te and Ga$_{1-x}$Mn$_{x}$As,
the 3$d$ transition-metal ions are substituted for the
cations of the host semiconductors. In Ga$_{1-x}$Mn$_{x}$As, the
ferromagnetism is induced by the hole carriers created by the substitution
of the divalent Mn ions for the trivalent cations. This ferromagnetism
is therefore called ``carrier-induced ferromagnetism''.
From the view point of applications, it is required to synthesize DMS with
a Curie temperature ($T_C$) above the room temperature. So far,
the $T_C$ of the III-V DMS has been mostly below the
room temperature. In recent theoretical studies, it has been
predicted that the ferromagnetic state is stable in 
Ga$_{1-x}$Mn$_{x}$N \cite{Sato, Mahadevan}
and that the ferromagnetism with a very high $T_C$ realized in
wide-gap systems such as $p$-type Ga$_{1-x}$Mn$_{x}$N \cite{Dietl}.
After the successful Mn doping into GaN \cite{Kuwabara, Zajac1},
several groups indeed reported that Ga$_{1-x}$Mn$_{x}$N
shows indication of ferromagnetism \cite{Sonoda, Overberg, Chen, Hashimoto, Yu}.
However, these results have been quite diverse between different
reports and the occurrence of ferromagnetism remains
controversial. Ando has reported that magnetic circular
dichroism (MCD) of Ga$_{1-x}$Mn$_{x}$N shows
a paramagnetic behavior \cite{Ando}.
It is also possible that Mn-N or Ga-Mn compounds are formed in
the crystal and contribute to the ferromagnetic 
properties \cite{Kuwabara, Dhar, Zajac2}.
It is therefore desirable
to characterize the electronic structure of Ga$_{1-x}$Mn$_{x}$N
to see whether there is an intrinsic possibility of high-temperature 
ferromagnetism in this system.

High-energy spectroscopic methods such as photoemission 
spectroscopy and soft x-ray absorption spectroscopy
are powerful techniques to investigate the electronic structure of solids.
In the studies of DMS, too, these techniques have played important
roles \cite{Jun, Wi}.
In this work, we have
investigated the electronic structure of Ga$_{1-x}$Mn$_{x}$N using
photoemission spectroscopy, soft x-ray absorption spectroscopy (XAS)
and subsequent configuration-interaction (CI) cluster-model
analysis. The CI approach is a useful tool to describe such
systems in which Coulomb interaction on the transition-metal
atom and strong hybridization between the transition-metal and
surrounding atoms coexist. The electronic structure parameters can
be estimated by analyzing the core-level and valence-band
photoemission spectra of DMS, enabling us to estimate
the magnitude of the interaction between the localized 3$d$ spins
and itinerant carriers.

\section{experimental}
The photoemission experiments were performed at BL-18A of Photon Factory,
High Energy Accelerator Research Organization (KEK). Photoelectrons
were collected using a VG CLAM hemispherical analyzer in the
angle-integrated mode. The total energy resolution including the
monochromator, the electron analyzer and the temperature broadening
was $\sim$200 meV as estimated from the Fermi edge of a metal. Core-level
x-ray photoemission (XPS) spectra were taken using a JEOL JPS-9200
hemispherical analyzer equipped with a Mg $K\alpha$ source
(h$\nu$ = 1253.6 eV) with the resolution of
$\sim$800 meV. Ghosts of the source due to the $K\alpha_{2}$, $K\alpha_{3,4}$
and $K\alpha_{5,6}$ lines
have been numerically subtracted. The photoemission spectra were
referenced to the Fermi edge of a metal in electrical contact with
the sample. 
All the measurements were made
in an ultra-high vacuum below 5$\times$10$^{-10}$ Torr at room temperature.
The x-ray absorption spectroscopy measurements were performed at the
soft x-ray beam line BL-23SU of SPring-8. Absorption
spectra were measured by the total electron yield method with the energy
resolution $E$/$\Delta$$E$ higher than 10000.
\begin{figure}[t]
\includegraphics[width=6.5cm]{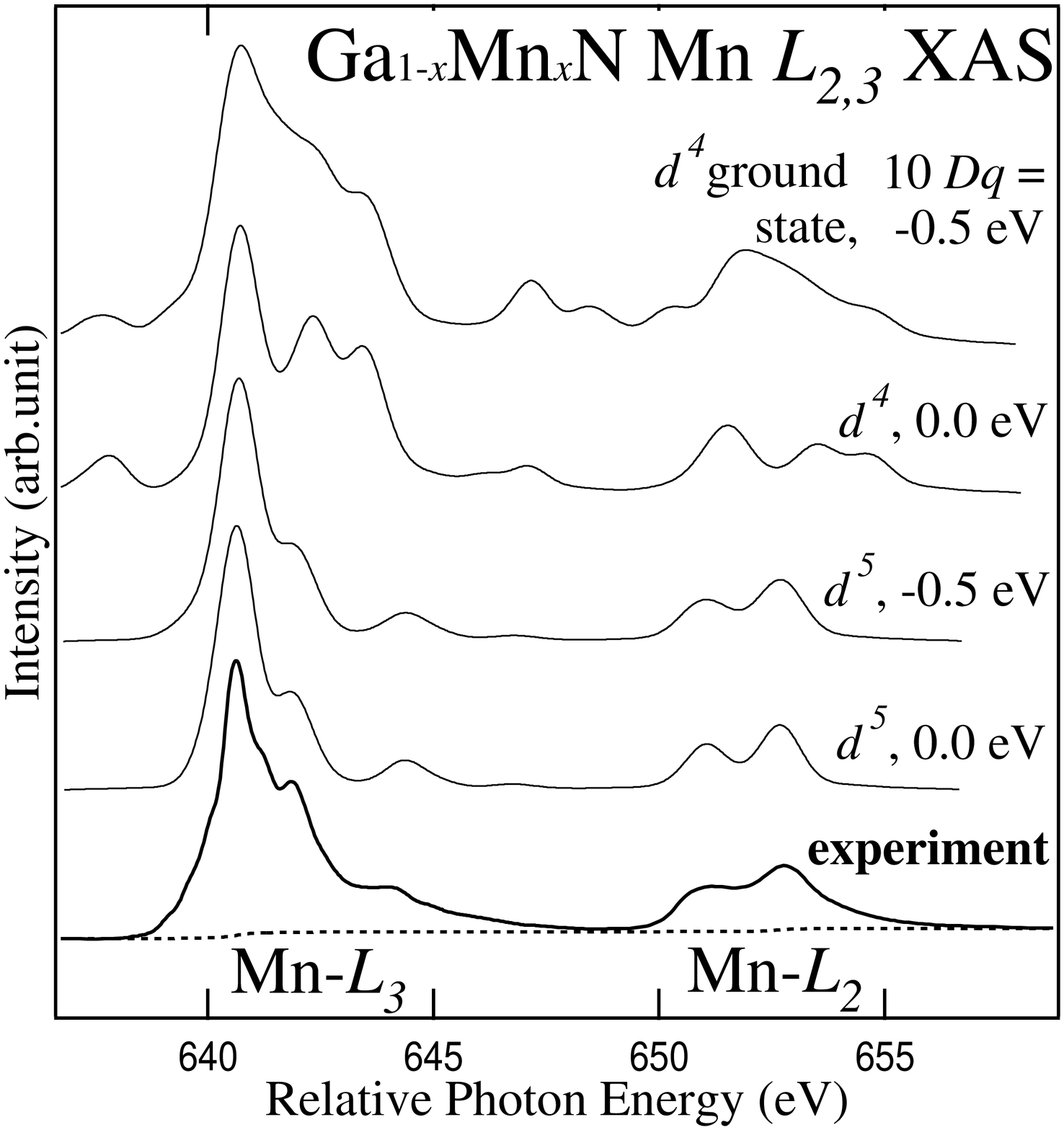}
\caption{Mn $L_{2,3}$-edge XAS of Ga$_{0.958}$Mn$_{0.042}$N. Calculations
for the $d^4$ and $d^5$ ground state in a tetrahedral crystal field (thin solid lines)
are compared with experimental data measured at 80 K (thick solid line)
with an arctangent type background
(dashed line). The line shapes of the experimental data are close to the calculated
spectra assuming the $d^5$ ground state with the crystal field $\sim$ -0.5 eV.}
\label{Mn2pXAS}
\end{figure}

Ga$_{1-x}$Mn$_{x}$N (0001) thin films with $x$ = 0.0, 0.02 and 0.042
were grown by molecular beam epitaxiy with an RF-plasma nitrogen
source and elemental sources of Ga and Mn on a sapphire
(0001) \cite{Kondo}. After nitridation of the substrate, an AlN
buffer layer (3 nm) was grown on the substrate at the substrate temperature
$T_s$ = 750 $^{\circ}$C followed by the growth of
a GaN buffer layer (100 nm) at $T_s$ = 700 $^{\circ}$C. On top of the GaN layer, Sn-doped $n$-type GaN
layer (110 nm) was grown to secure the conduction of the sample.
Finally, a Ga$_{1-x}$Mn$_{x}$N epitaxial layer (110 nm) was deposited
on top of it at the substrate temperature of 550 $^{\circ}$C. All the
samples thus prepared were paramagnetic from room temperature down to 4 K. For
surface cleaning, we made N$_{2}^{+}$ ion sputtering, which 
compensates the loss of N atoms, followed by
annealing up to 500$^{\circ}$C
because Ar$^{+}$ ion preferentially sputters N atoms 
and may induce excess Ga atoms to form Ga clusters on the surface \cite{Lai}.
We have confirmed that the photoemission
spectra did not change by annealing.
The cleanliness of the surface was checked by
low-energy electron diffraction (LEED)
and core-level XPS. The O 1$s$ and C 1$s$ core-level peaks
were diminished below the detectability limit by repeating N$_{2}^{+}$
ion-sputtering and annealing, and clear a LEED pattern was obtained,
reflecting ordered clean surfaces.
\begin{figure}[t]
\includegraphics[width=6.5cm]{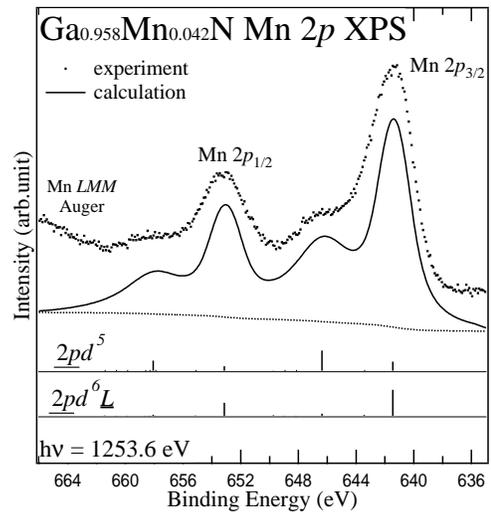}
\caption{
Mn 2$p$ core-level XPS of
Ga$_{0.958}$Mn$_{0.042}$N (dots) and its CI cluster-model analysis
(solid line). In the bottom, the calculated spectrum has been
decomposed into $\underline{2p}d^{5}$
and $\underline{2p}d^{6}\underline{L}$ final-state configurations.
The calculations well reproduce the satellite
structures with the electronic structure parameter
$\Delta$ = 4.0 $\pm$ 1.0, $U$ = 6.0 $\pm$ 1.0 and ($pd\sigma$) = -1.5 $\pm$ 0.1 eV.}
\label{Mn2pwithCI}
\end{figure}

\section{results and discussion}
Figure \ref{Mn2pXAS} shows an XAS spectrum at the Mn $L$-edge compared with atomic multiplet calculations for the $d^{4}$ and $d^{5}$ ground states.
The rich structures of the observed spectrum show a multiplet splitting and are
typical for localized 3$d$ states. 
Comparing the experimental line shape
with the calculation, one can obtain the information about
the valence and the crystal field of Mn in Ga$_{1-x}$Mn$_{x}$N.
From Fig. \ref{Mn2pXAS}, one can see that the
calculation assuming the $d^{5}$ ground state with the tetrahedral-crystal
field 10$Dq$ $\sim$ -0.5 eV well represents the
experimental spectrum. The negative 10$Dq$ means a crystal field due to the
tetrahedrally coordinating anions, while positive 10$Dq$ means a crystal field
with octahedral symmetry such as the interstitial position 
in the wurtzite structure \cite{Popovic}. 
This leads us to conclude that most of Mn atoms
in the sample used in the present measurements are in
the $d^{5}$ state, namely,
the valence of Mn is divalent with
the total spin $S$ = 5/2 and in the tetrahedral crystal field.
A similar line shape of the absorption spectra of Ga$_{1-x}$Mn$_{x}$N 
and a similar conclusion about the valence 
state have been reported
\cite{Edmonds}. This is also consistent with
the previous report about the valence state of Mn in GaN 
based on the Curie-Weiss behavior of the magnetic susceptibility
data \cite{Kuwabara}.
On the other hand,
the valence state of Mn in non-doped GaN has been reported to be
trivalent based on electron spin resonance and optical
absorption \cite{Graf1}. The reason for this difference is not clear at present,
but it has also been reported that Mn is divalent 
when electrons are doped \cite{Graf2} whereas Mn is tetravalent when holes are doped \cite{Korotkov}. 
It has also been reported that the charge transfer occurs 
across adjacent layer in Ga$_{1-x}$Mn$_{x}$N/p-GaN:Mg and 
Ga$_{1-x}$Mn$_{x}$N/n-GaN:Si heterostructures \cite{Arkun}. 
In that report, it has been 
indicated that a large number of electrons (holes) are 
transferred from adjacent n-type (p-type) layer into the sample.
The difference thus may be explained by the scenario of the charge 
transfer from Sn-doped $n$-type GaN to Ga$_{1-x}$Mn$_{x}$N 
(i.e., electron doping into Ga$_{1-x}$Mn$_{x}$N).
\begin{figure}[tbp]
\includegraphics[width=4.2cm]{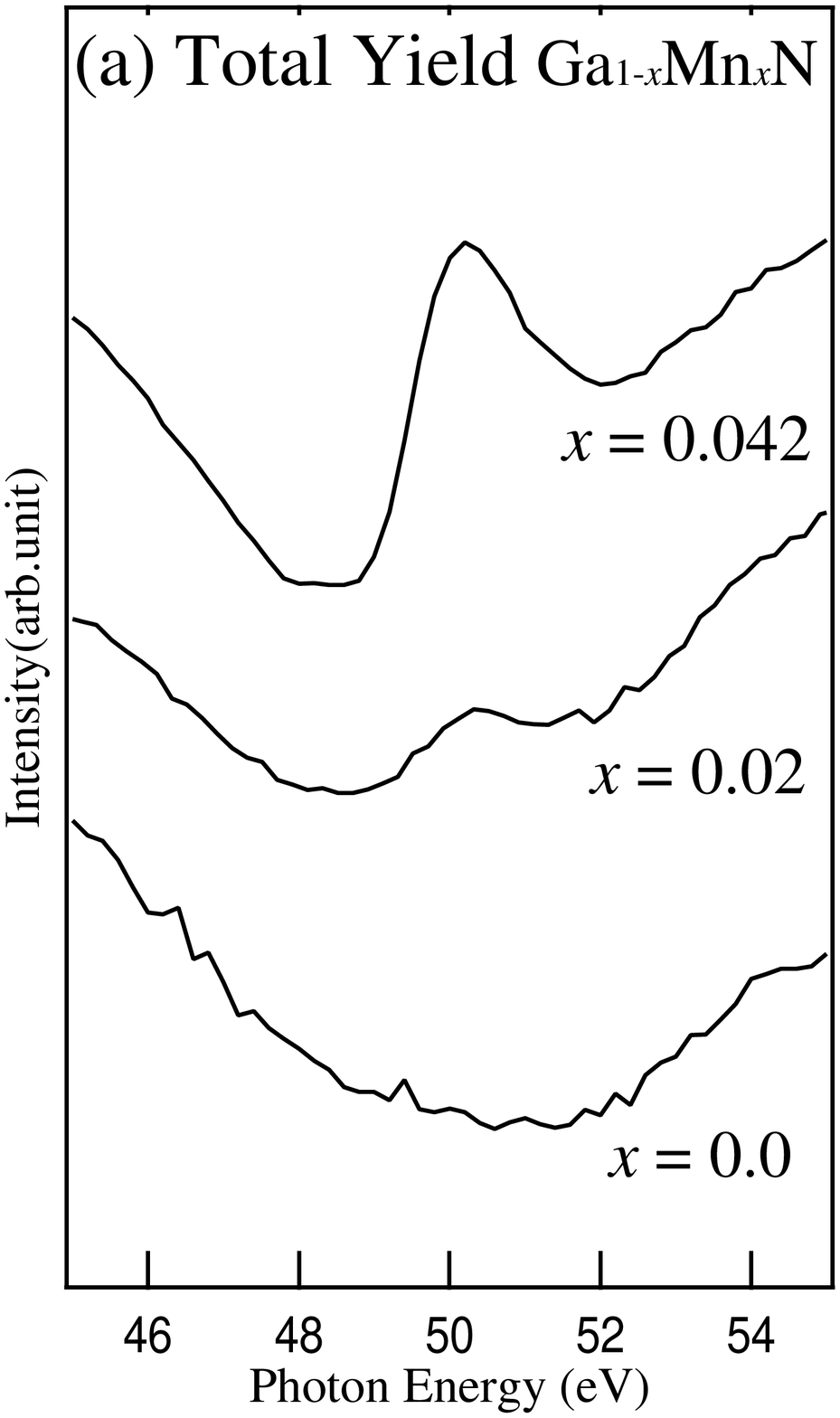}
\includegraphics[width=4.2cm]{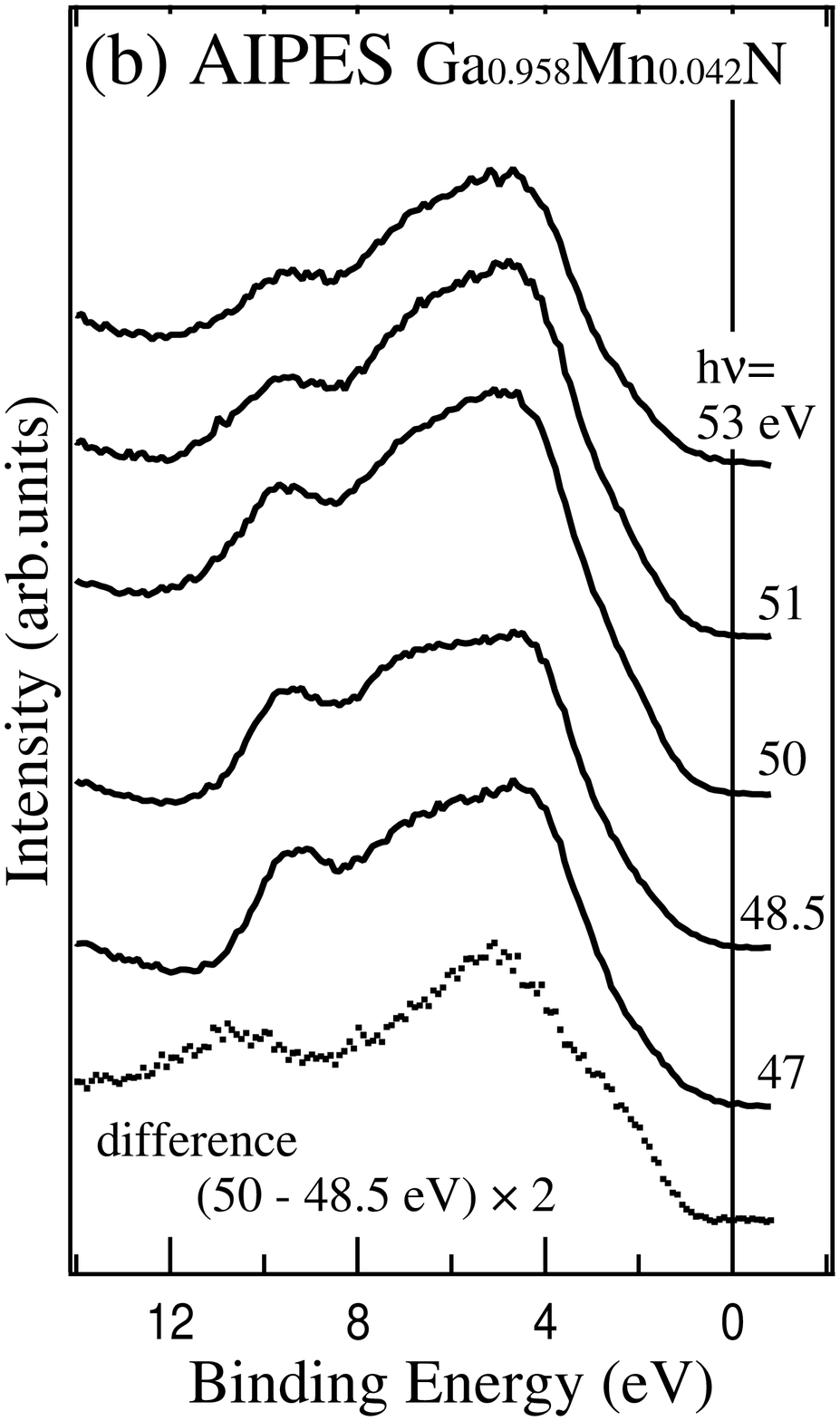}
\caption{(a)Mn 3$p$-3$d$ absorption spectra
of Ga$_{1-x}$Mn$_{x}$N with $x$ = 0.0, 0.02
and 0.04 recorded by the total electron yield
method. (b) Photoemission spectra of
Ga$_{0.958}$Mn$_{0.042}$N for various photon energies around
the Mn 3$p$-3$d$ core excitation threshold. The difference curve
at the bottom represents the Mn 3$d$ PDOS.}
\label{445AIPES}
\end{figure}

Figure \ref{Mn2pwithCI} shows the Mn 2$p$ core-level photoemission spectra
of the $x$ = 0.042 sample. The broad peak at 667 eV is due to overlapping Mn
$L_{2,3}M_{2,3}M_{4,5}$ Auger emission. The spectra show a spin-orbit
doublet, each component of which shows a satellite structure on
the higher binding energy side separated by $\sim$ 5 eV. The presence
of the satellite structure indicates strong Coulomb interaction
among the 3$d$ electrons and strong hybridization between the 3$d$
electrons and the host valence electrons.
\begin{figure}[tbp]
\includegraphics[width=7cm]{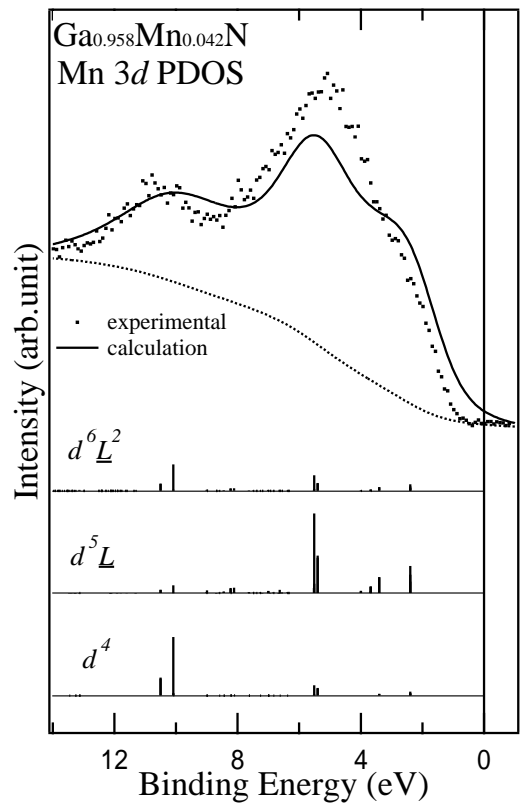}
\caption{
Mn 3$d$ PDOS of Ga$_{0.958}$Mn$_{0.042}$N obtained by
RPES (dots) and its CI cluster-model
analysis (solid line). At the bottom, the calculated spectrum
has been decomposed into $d^{5}$, $d^{6}\underline{L}$
and $d^{7}\underline{L}^{2}$ final-state configurations.}
\label{Mn3dwithCI}
\end{figure}

In order to analyze the satellite structure and to obtain the electronic structure
parameters, atomic multiplet theory has to be extended to the CI cluster
model in order to take into account charge transfer effects. We have
employed a tetrahedral MnN$_4$
cluster with the central Mn atom. Here, the small distortion from the
tetrahedron in the wurtzite structure is ignored. In the CI picture,
the wave function of the ground state is given by a linear combination
of $d^{5}$, $d^{6}\underline{L}$, $d^{7}\underline{L}^{2}$, $\cdots$,
where $\underline{L}$ denotes a hole in the ligand $p$ orbitals.
The final states of Mn 2$p$ photoemission spectra are given
by linear combinations of $\underline{2p}d^{5}$,
$\underline{2p}d^{6}\underline{L}$, $\cdots$,
$\underline{2p}d^{10}\underline{L}^{5}$, where $\underline{2p}$ stands
for a Mn 2$p$ core hole . The anion-to-3$d$ orbital charge-transfer energy
is $\Delta$ =$E(d^{n+1}\underline{L})$ - $E(d^{n})$,
and the 3$d$-3$d$ Coulomb interaction energy is defined by
$U$ = $E(d^{n+1})$ + $E(d^{n-1})$ - 2$E(d^{n})$, where
$E(d^{l}\underline{L}^{k}$) denoted the center of gravity of the
$d^{l}\underline{L}^{k}$ multiplet. The average attractive Coulomb energy
$Q$ between the Mn 3$d$ electron and the Mn 2$p$ core hole has been fixed at
$U$/$Q$ = 0.8. The transfer integrals between the 3$d$ and ligand $p$
orbitals are given by Slater-Koster parameters ($pd\sigma$) and ($pd\pi$)
assuming the relationship ($pd\sigma$)/($pd\pi$) = -2.16 \cite{Harrison}.
The calculated spectrum has been broadened with a Gaussian and a Lorentzian.
We have assumed that the valence of the Mn is divalent according to the XAS result.
We have also ignored additional holes which enter into the top of
the valence band of GaN because the carrier concentration in the present
samples is negligibly small, judged from the very high resistivity,
presumably due to charge compensation. Thus, the satellite
structures are well reproduced with parameter values
$\Delta$ = 4.0 $\pm$ 1.0, $U$ = 6.0 $\pm$ 1.0 and ($pd\sigma$) = -1.5 $\pm$ 0.1 eV.

We have also investigated the valence band using resonant photoemission
spectroscopy (RPES) to extract the Mn 3$d$ partial density of states (PDOS).
In RPES, because Mn 3$p$-to-3$d$ absorption occurs at photon
energies above 50 eV, interference between the normal
photoemission and 3$p$-to-3$d$ transition followed by a
3$p$-3$d$-3$d$ super-Coster-Kr\"onig decay generates a resonance
enhancement of the Mn 3$d$-drived photoemission. 
Figure \ref{445AIPES} (a)
shows the absorption spectra near the Mn 3$p$-3$d$ core excitation
threshold of Ga$_{1-x}$Mn$_{x}$N measured by the total electron yield method.
In the Fig. \ref{445AIPES} (a), one can see that with increasing $x$,
a peak appears at 50 eV and grows in intensity, representing the
Mn 3$p$-to-3$d$ absorption. From these spectra, on-resonance
and off-resonance photon energies are found to be 50 and 48.5 eV, respectively.
Figure \ref{445AIPES} (b) shows the valence-band spectra of the
$x$ = 0.042 sample taken at various photon energies in the Mn
3$p$-to-3$d$ core excitation region. The intensities have been
normalized to the photon flux. All binding energies are referenced
to the Fermi level ($E_{F}$). In going from h$\nu$ = 47 to 50 eV,
one can see that the peak at the binding energy of 5 eV grows
in intensity. By subtracting the off-resonant spectrum from the
on-resonant one, we obtained the Mn 3$d$ derived spectrum as shown
in the bottom panel of Fig. \ref{445AIPES} (b). For the subtraction, the photon
energy dependence of the photoionization-cross section of 
the N 2$p$ atomic orbital has
been considered. The difference spectrum that corresponds to the Mn 3$d$
PDOS reveals a peak at $E_{B}$ = 5 eV and a shoulder at $E_{B}$ = 2 eV.
The shoulder is located at $\sim$ 0.5 eV above VBM and well below the Fermi 
level due to charge compensation.
The satellite also appears at $E_{B}$ = 9 - 13 eV, at a higher
binding energy than that of Ga$_{1-x}$Mn$_{x}$As \cite{Jun}. 
\begin{table}[tbp]
\begin{center}
\caption{Electronic structure parameters in units of eV for substitutional
Mn impurities in semiconductors and estimated $p$-$d$ exchange
constant $N\beta$ for Mn$^{2+}$.} 
\begin{ruledtabular}
\begin{tabular}{lccccc}
Material            & $\Delta$ & $U$ & $(pd\sigma)$ & $N\beta$ & Reference \\
\hline
Ga$_{1-x}$Mn$_{x}$N  & 4.0      & 6.0 & -1.5         & -1.6     & This work   \\
Ga$_{1-x}$Mn$_{x}$As & 1.5      & 3.5 & -1.0         & -1.0     & 14 \\
In$_{1-x}$Mn$_{x}$As & 1.5      & 3.5 & -0.8         & -0.7     & 27\\
Zn$_{1-x}$Mn$_{x}$O  & 6.5      & 5.2 & -1.6 	    & -2.7 & 28\\
Zn$_{1-x}$Mn$_{x}$S  & 3.0      & 4.0 & -1.3         & -1.3     & 28\\
Zn$_{1-x}$Mn$_{x}$Se & 2.0      & 4.0 & -1.1         & -1.0     & 28\\
Zn$_{1-x}$Mn$_{x}$Te & 1.5      & 4.0 & -1.0         & -0.9     &\ 28
\label{Table}
\end{tabular}
\end{ruledtabular}
\end{center}
\end{table}
\begin{figure}[h]
\includegraphics[width=4.2cm]{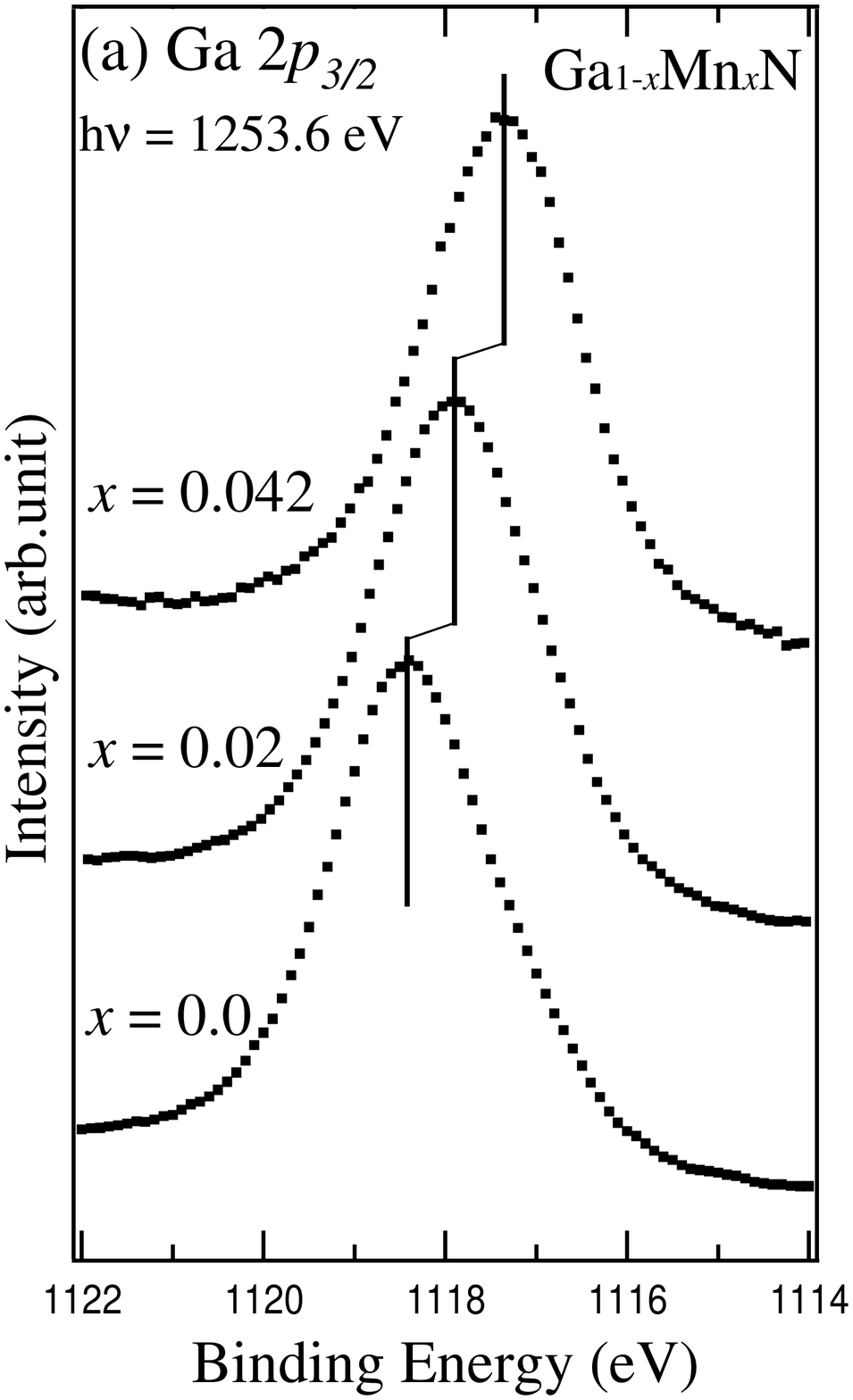}
\includegraphics[width=4.2cm]{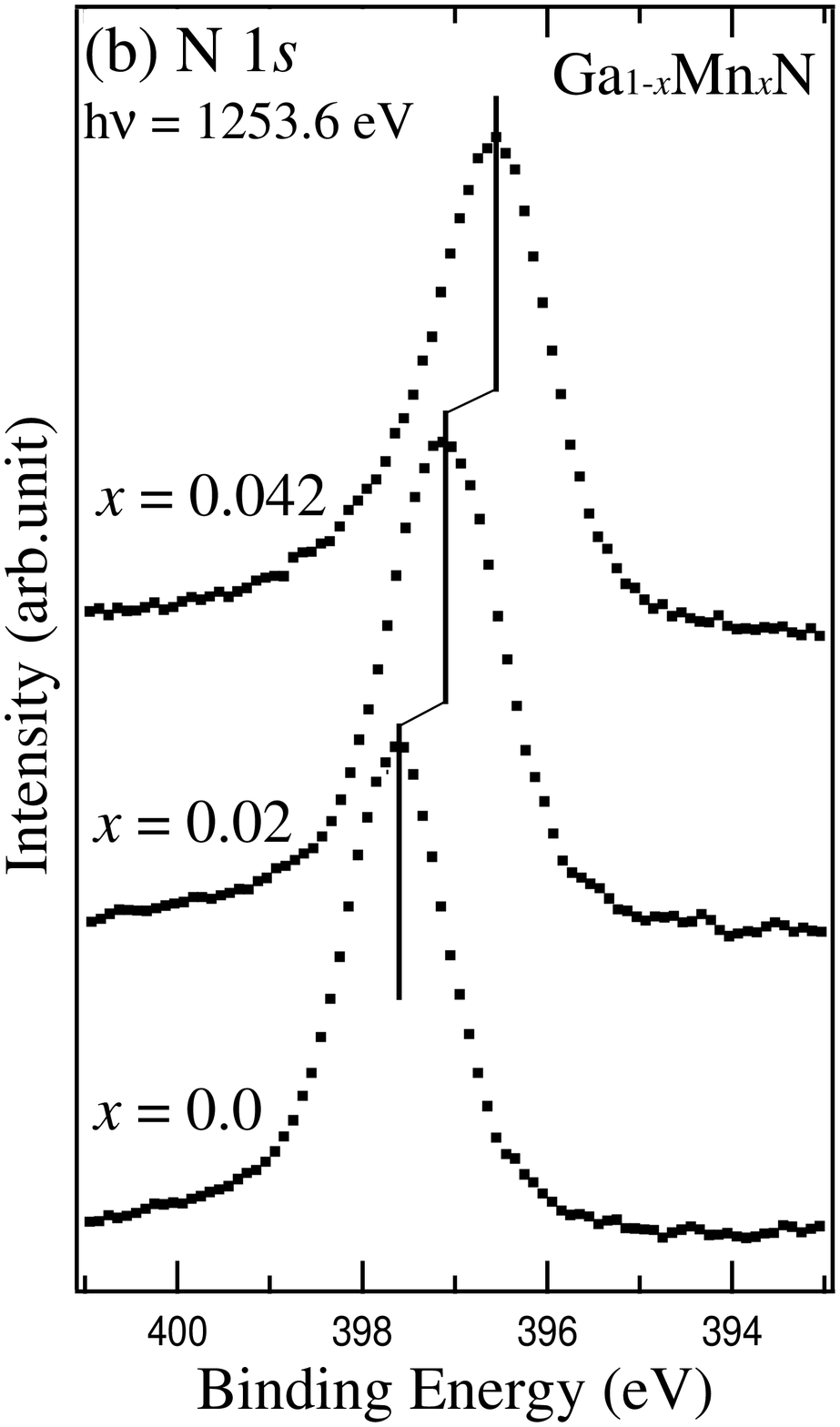}
\caption{Composition dependence of the core-level 
photoemission spectra of Ga$_{1-x}$Mn$_{x}$N. (a)Ga 2$p_{3/2}$ core level. 
(b)N 1$s$ core level. }
\label{XPS}
\end{figure}

CI cluster-model calculation has also been used to analyze the Mn 3$d$ PDOS
as shown in Fig. \ref{Mn3dwithCI}.
One can see that the calculation well explains not only
the main structure but also the satellite which could not be explained by the band-structure calculation \cite{Sato}, with identical
parameters as in the case of Mn 2$p$ core-level within the error bars,
$\Delta$ = 4.0 $\pm$ 1.0, $U$ = 6.0 $\pm$ 1.0 and ($pd\sigma$) = -1.5 $\pm$ 0.1 eV.
As shown in the bottom panels of
Fig. \ref{Mn3dwithCI}, the main peak largely consists of
$d^{5}\underline{L}$ final states and the satellite of $d^{4}$ final states
because of $\Delta \textless U$.
In Table \ref{Table}, the parameter values for Ga$_{1-x}$Mn$_{x}$N 
are compared with those of other DMS.
$\Delta$, $U$, and ($pd\sigma$) for Ga$_{1-x}$Mn$_{x}$N
are large comparing with other III-V and II-VI DMS, except for
Zn$_{1-x}$Mn$_{x}$O for which $\Delta$, $U$ and ($pd\sigma$)
are large. The charge-transfer energy $\Delta$ of Ga$_{1-x}$Mn$_{x}$N
is large because of the high electronegativity of the anion.
The one-electron transfer integral ($pd\sigma$) is large because
of the small distance between Mn and the anion.
The large value of $U$ may be attributed to the low polarizability of N atom.

Using the $\Delta$, $U$ and ($pd\sigma$) values thus obtained, one can
estimate the $p$-$d$ exchange constant $N\beta$ for the Mn$^{2+}$
ion in the semiconductor host in the second order of perturbation with
respect the hybridization term:
\begin{displaymath}
\label{eq:Nb}
N\beta = - \frac{16}{S}\Biggl( \frac{1}{U_{\textrm {eff}}-\delta_{\textrm {eff}}}+\frac{1}
{\delta_{\textrm {eff}}}\Biggr) 
\Biggl(\frac{1}{3}(pd\sigma) - 
 \frac{2\sqrt{3}}{9}(pd\pi)\Biggr)^{2},
\end{displaymath}
where $\delta_{\textrm {eff}}$ is defined 
as $\delta_{\textrm {eff}}$ = $\Delta_{\textrm {eff}}$ - $W_{V}$/2.
The valence band width $W_{V}$ has been fixed at 3 eV because
the upper 3 eV of the host
valence band primarily contributes to hybridization term although
the total width of the host valance band is 5-6 eV \cite{band}.
The charge-transfer energy $\Delta_{\textrm {eff}}$ and the 3$d$-3$d$ Coulomb
interaction energy $U_{\textrm {eff}}$ are defined with respect to the
lowest term of each multiplet.
Using Racah parameters $B$ and $C$ of the free
Mn$^{2+}$ ion values ($B$ = 0.119 eV and $C$ = 0.412 eV) \cite{Racah}, $\Delta_{\textrm {eff}}$
and $U_{\textrm {eff}}$ are given by $\Delta_{\textrm {eff}}$ = $\Delta$ + (70$B$ - 35$C$)/9 + 7$C$
and $U_{\textrm {eff}}$ = $U$ + (14$B$ - 7$C$)/9 +14$B$ + 7$C$. The local spin $S$
is 5/2 for Mn$^{2+}$ ($d^{5}$). The value of $N\beta$ thus estimated is -1.6
$\pm$ 0.3 eV
and is larger than that of other III-V DMS (see Table 1).
While the differences between $\Delta$ and $U$, which contribute to
the denominator of the equation, is similar for
Ga$_{1-x}$Mn$_{x}$N, Ga$_{1-x}$Mn$_{x}$As and In$_{1-x}$Mn$_{x}$As,
($pd\sigma$) is large in Ga$_{1-x}$Mn$_{x}$N, giving rise to
the large $N\beta$ in Ga$_{1-x}$Mn$_{x}$N. 
\begin{figure}[h]
\includegraphics[width=4.2cm]{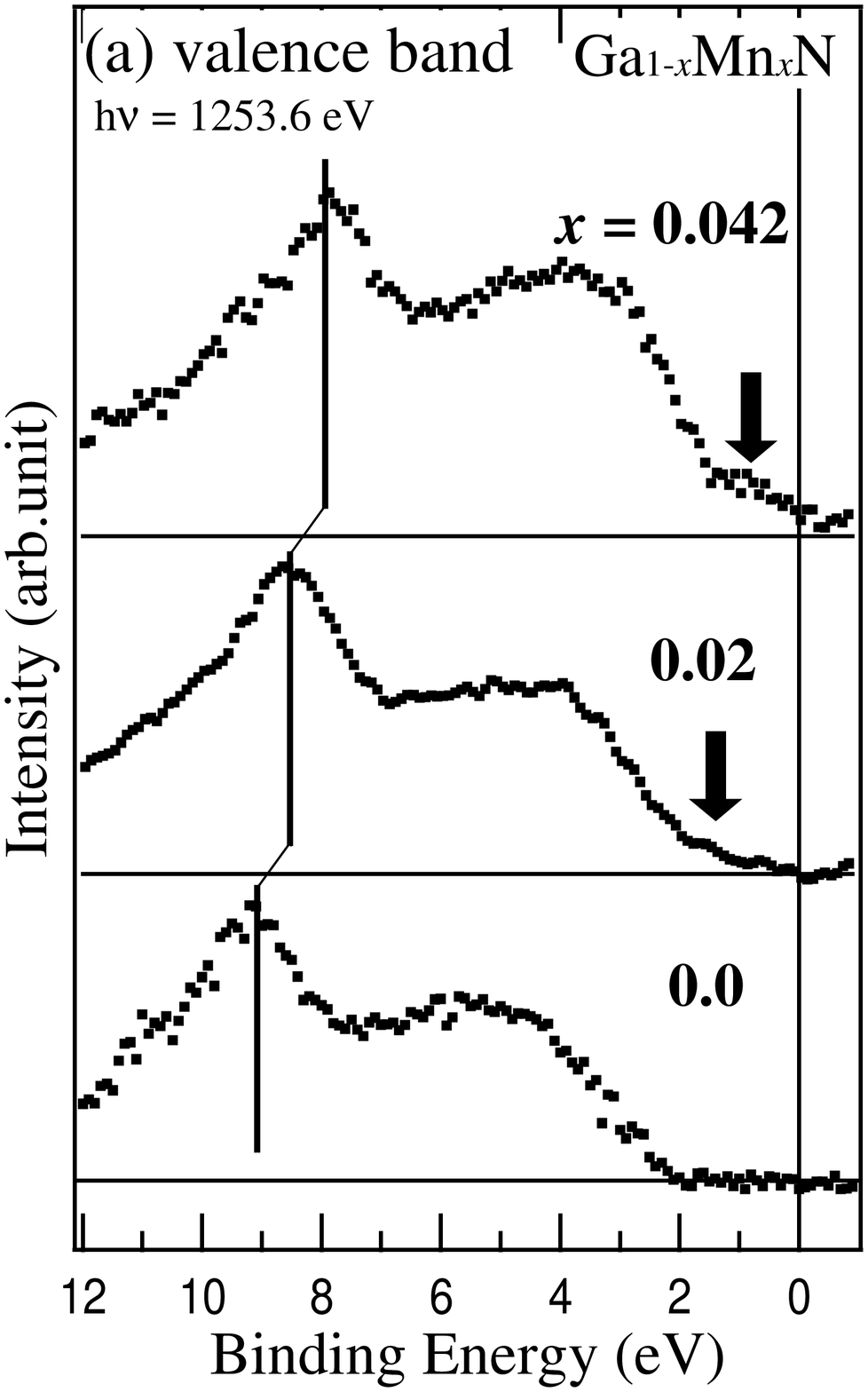}
\includegraphics[width=4.2cm]{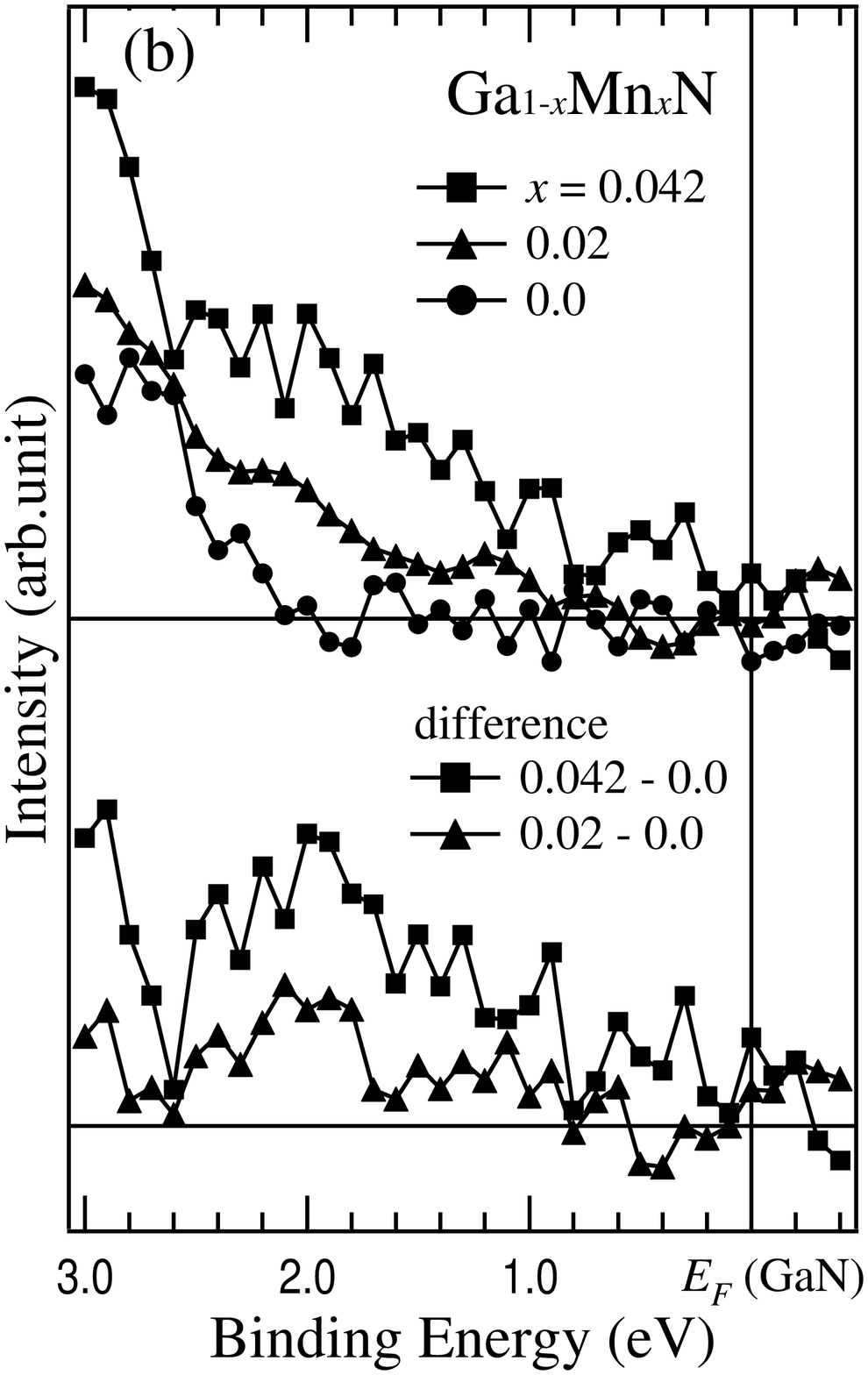}
\caption{Composition dependence of the valence-band photoemission spectra of 
Ga$_{1-x}$Mn$_{x}$N. (b) Plots on an expanded scale near the 
valence-band maximum, where the spectra of 
$x$ = 0.02 and 0.042 have been shifted to higher binding energy compared
with $x$ = 0.0. Difference curves are shown in the bottom panel.}
\label{vXPS}
\end{figure}

\begin{figure}[h]
\includegraphics[width=7.5cm]{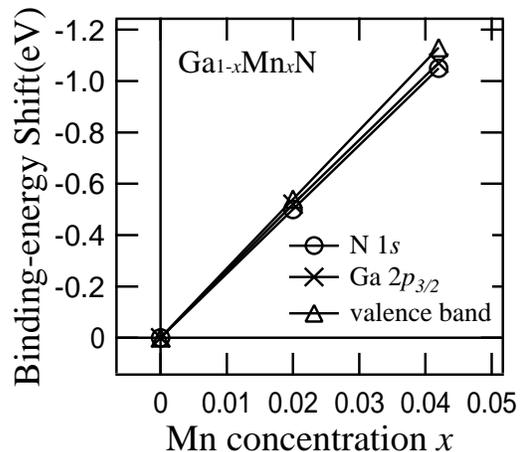}
\caption{Energy shifts of the core levels and the valence band as a function
of Mn concentration for Ga$_{1-x}$Mn$_{x}$N.}
\label{Shift}
\end{figure}

Finally, we show the Mn concentration dependence of the electronic
structure of Ga$_{1-x}$Mn$_{x}$N. The Ga 2$p_{3/2}$ and N 1$s$
core-level photoemission spectra in Fig. \ref{XPS}(a) and (b)
clearly show 
shifts towards lower binding energies with increasing Mn concentration.
Nearly the same amount of shifts were also observed in the valence-band spectra,
as shown in Fig. \ref{vXPS}(a). Note that these shifts cannot be due to charging effect
because the electrical resistivity increases with Mn doping. 
Therefore, we conclude that the Fermi level is shifted
downward with Mn doping, that is, the doped Mn atoms supply holes 
and compensate the electron carriers. 
From $x$ = 0 to 0.02 to 0.042, the amount of these shifts 
are 0.5 and 1.05 eV for N 1$s$, 0.52 and 1.07 eV for Ga 2$p_{3/2}$, 0.54 and 1.13 eV for the valence band. The shifts are linear with $x$ as shown in Fig. \ref{Shift}. 
In the valence-band spectra
[Fig. \ref{vXPS} (a)], too, one can see that upon doping, new states are
created above the VBM as shown by arrows.
The new states are consistent with the result of a recent photoemission study
using hard x-rays \cite{Kim}. 
The spectra near the valence-band maximum are shown in Fig. \ref{vXPS} (b) 
and difference spectra from undoped GaN are shown in the bottom panel. 
Here, prior to the subtraction, the spectra of the $x$ = 0.02 
and 0.042 have been shifted to higher binding energy so that the peak at $E_B$
$\sim$8eV is aligned.
in order to deduce changes in the DOS induced by Mn doping.
The new states are located at $\sim$ 0.5 eV above the VBM and are 
increased in intensity with increasing 
Mn concentration.
This is considered as due to the appearance of 
Mn 3$d$ character because in XPS the relative photoionization-cross
section of Mn 3$d$ to N 2$p$ is as large as 10. 
If the holes doped through the
Ga $\to$ Mn substitution were not compensated, the acceptor level formed
above the VBM would remain empty (i.e., above $E_F$) and would not be observed
in the photoemission spectra. The new states above VBM are 
therefore considered to be the
acceptor levels occupied by compensating electrons.

One of the reasons why ferromagnetism does not occur in 
Ga$_{1-x}$Mn$_{x}$N samples used in this study in spite of the large
$p$-$d$ exchange constant $N\beta$ may be the lack of a sufficient
number of hole 
carriers which mediate ferromagnetic coupling between Mn ions. The new
states above the VBM seen in Fig. \ref{vXPS}(b) are consistent with
this charge compensation picture. This suggests that if the compensation is reduced
in some way and sufficient amount of holes are doped into the system,
strong ferromagnetism would be expected.

\section{conclusion}
In conclusion, we have investigated the electronic structure of
the diluted magnetic semiconductor Ga$_{1-x}$Mn$_{x}$N using
high-energy spectroscopy. It is confirmed that Mn in Ga$_{1-x}$Mn$_{x}$N
is divalent. We have treated the electronic structure as a many-electron 
system using CI analysis. The CI analysis of photoemission 
spectra on a MnN$_{4}$ cluster
model reveals that the magnitude of $p$-$d$ exchange constant $N\beta$ in
Ga$_{1-x}$Mn$_{x}$N is much larger than that in Ga$_{1-x}$Mn$_{x}$As.
This implies the possibility that $p$-$d$ exchange 
contributes to realize the room temperature
ferromagnetism.
Mn substitution indeed introduces holes into the system although most
of them are compensated.

We thank T. Okuda, A. Harasawa and T. Kinoshita for their valuable
technical support. This work was supported by a Grant-in-Aid for
Scientific Research in Priority Area gSemiconductor Spintronicsh (14076209)
from the Ministry of Education, Culture, Sports, Science and Technology,
Japan. One of us (HO) acknowledges support from German Academy Exchange Service.
The experiment at Photon Factory was approved by the Photon Factory
Program Advisory Committee (Proposal No. 2002G027).

\end{document}